\documentclass{article}

\usepackage{amssymb,amsfonts,amsmath}
\usepackage{cite,enumerate,float}
\usepackage{color}
\usepackage{tikz}
\usetikzlibrary{arrows,snakes,backgrounds}
\usepackage{ytableau}
\usepackage[vcentermath]{youngtab}

\def\be{\begin{eqnarray}}
\def\ee{\end{eqnarray}}
\def\nn{\nonumber}

\def\p{\partial}

\def\Tr{{\rm Tr}\,}

\definecolor{red}{rgb}{1,0,0}
\definecolor{orange}{rgb}{1,0.5,0}
\definecolor{violet}{rgb}{0.7,0,1}

\hoffset= - 1.0in         % switch off for draft style
\begin{document}

\title{\vspace{1.5cm}\bf
On KP-integrable skew Hurwitz $\tau$-functions\\ and their $\beta$-deformations}

\author{
A. Mironov$^{b,c,d,}$\footnote{mironov@lpi.ru,mironov@itep.ru},
V. Mishnyakov$^{a,b,}$\footnote{mishnyakovvv@gmail.com},
A. Morozov$^{a,c,d,}$\footnote{morozov@itep.ru},
A. Popolitov$^{a,c,d,}$\footnote{popolit@gmail.com},
Wei-Zhong Zhao$^{e,}$\footnote{zhaowz@cnu.edu.cn}
}

\date{ }

\maketitle

\vspace{-6.5cm}

\begin{center}
\hfill FIAN/TD-02/23\\
\hfill IITP/TH-02/23\\
\hfill ITEP/TH-02/23\\
\hfill MIPT/TH-02/23
\end{center}

\vspace{4.5cm}

\begin{center}
$^a$ {\small {\it MIPT, Dolgoprudny, 141701, Russia}}\\
$^b$ {\small {\it Lebedev Physics Institute, Moscow 119991, Russia}}\\
$^c$ {\small {\it ITEP, Moscow 117218, Russia}}\\
$^d$ {\small {\it Institute for Information Transmission Problems, Moscow 127994, Russia}}\\
$^e${\em School of Mathematical Sciences, Capital Normal University,
Beijing 100048, China} \\
\end{center}

\vspace{.0cm}

\begin{abstract}
We extend the old formalism of cut-and-join operators in the theory of Hurwitz $\tau$-functions to description of
a wide family of KP-integrable {\it skew} Hurwitz $\tau$-functions, which include, in particular, the newly discovered
interpolating WLZZ models.  Recently, the simplest of them was related to a superintegrable two-matrix model with two potentials and one external matrix field. Now we provide detailed proofs, and a generalization to a multi-matrix representation, and propose the $\beta$-deformation of the matrix model as well. The general interpolating WLZZ model is generated by a $W$-representation given by a sum of operators from a one-parametric commutative sub-family (a commutative subalgebra of $w_\infty$). Different commutative families are related by cut-and-join rotations. Two of these sub-families (`vertical' and `45-degree')
turn out to be nothing but the trigonometric and rational Calogero-Sutherland Hamiltonians, the `horizontal' family is
represented by simple derivatives. Other families require an additional analysis.
\end{abstract}

\section{Introduction}

Matrix models are the simplest representatives of the universality classes of
non-perturbative partition functions, which exhibit their peculiar properties,
well shadowed in perturbative Feynman diagram technique and in other approaches
based on the standard multi-dimensional quantum field theory.
These properties include high hidden symmetries, integrability, superintegability
and $W$-representations, which express partition functions through the action
of generalized cut-and-join operators on vacuum states.
By now, they are well-studied and understood in particular models,
and the new challenge is to classify and unite different models in a common entity,
as a step to a formulation of non-perturbative string theory.

Recently a big step was made in this direction by the suggestion of WLZZ models \cite{China},
which appeared to be building blocks of a two-matrix model in the external matrix field \cite{MMMPWZ}.
In this paper, in particular, we provide technical details behind this general claim made in
\cite{MMMPWZ}. They are based on using peculiar cut-and-join rotation operators  ${\cal O}$
and on an extension of the old formalism of \cite{AMMN} to
{\it skew} Hurwitz partition functions, naturally expanded in {\it skew} rather than ordinary
Schur functions, and exploit the Littlewood-Richardson decompositions of ones into the others.
These decompositions get a little non-trivial after the $\beta$-deformation
from the Schur functions to the Jack polynomials, which we also explain.

\bigskip

\paragraph{Skew Hurwitz partition functions.} Generic Hurwitz partition functions depending on two sets of time variables have the form \cite{MMN1},
\be\label{Hur}
Z_H(u_\Delta;g,p)=\sum_RS_R\{g_k\}S_R\{p_k\}\exp\left(\sum_\Delta u_\Delta\psi_R(\Delta)\right)
\ee
$S_R\{p_k\}$ is the Schur function labelled by the partition $R$, which is a symmetric function of $x_i$ considered as a graded polynomial of the power sums $p_k=\sum_ix_i^k$, $\Delta$ is a partition, and $\psi_R(\Delta)$ is the character of symmetric group $S_{|R|}$, $|R|$ being the size of partition \cite{Mac,F}. This partition function is generated by generalized cut-and-join operators \cite{MMN1}, and is not integrable in the KP/Toda sense \cite{DJKM,UT}, it becomes integrable only with a special choice of the coefficients $u_\Delta$:
\be\label{1}
\exp\left(\sum_\Delta u_\Delta\psi_R(\Delta)\right)\longrightarrow\exp\left(\sum_k u_kC^{(k)}_R\right)
\ee
where $C^{(k)}_R$ are eigenvalues of specially chosen $k$th Casimir operators  \cite{GKM2,MMN1,Ok,AMMN0}\footnote{They can be also associated with values of characters on the completed cycles \cite{Ok,Lando}.}. Equivalently,
one can choose \cite{OS,AMMN}
\be\label{2}
\exp\left(\sum_\Delta u_\Delta\psi_R(\Delta)\right)\longrightarrow \prod_{i,j\in R}f(j-i)
\ee
with an arbitrary function $f(x)$. KP-integrable Hurwitz partition functions of this kind are called hypergeometric $\tau$-functions of the KP hierarchy \cite{OS}.

In this paper, we propose a generalization of the Hurwitz partition functions (\ref{Hur}) to the skew Hurwitz partition functions,
\be
Z_H(u_\Delta,v_{\Delta'};g,p)=\sum_RS_{R/Q}\{\bar p_k\}S_R\{g_k\}S_Q\{p_k\}\exp\left(\sum_\Delta u_\Delta\psi_R(\Delta)
-\sum_{\Delta'} v_{\Delta'}\psi_Q(\Delta')\right)
\ee
where $S_{R/Q}$ is the skew Schur function \cite{Mac}. They are also integrable only with choices analogous to (\ref{1})-(\ref{2}) and are called skew hypergeometric $\tau$-functions \cite{MMMPWZ}.

In \cite{AMMN}, it was suggested to choose the function $f(x)$ in (\ref{2})
\be
f(x)=\prod_i^n(u_i+x)
\ee
This choice has many applications, from generating numbers of isomorphism classes of the Belyi pairs (Grothendieck’s dessins d’enfant) \cite{Gr} to the Itzykson-Zuber integral \cite{IZ}.

In this paper, as an extension of results of \cite{AMMN}, we consider an extension of this choice to the skew Hurwitz partition functions. Thus, we consider a partition functions of the form
\be\label{Pf}
\mathfrak{Z}_{(n,m)}(\{u_i\},\{v_j\};\bar p,g,p)=\sum_{R,Q}{\prod_{i=1}^n\xi_R(u_i)\over\prod_{j=1}^m\xi_Q(v_j)}S_{R/Q}\{\bar p_k\}S_R\{g_k\}S_Q\{p_k\}
\ee
where
\be
\xi_R(u)=\dfrac{S_R\{p_k=u\}}{S_R\{\delta_{k,1}\}}=\prod_{i,j\in R}(u+j-i)
\ee

The partition function $\mathfrak{Z}_{(n,m)}$ depends on $n$ parameters $u_i$, $m$ parameters $v_i$ and three infinite sets of time variables $\{g_k\}$, $\{p_k\}$, and $\{\bar p_k\}$.
The interpolating matrix model of \cite{MMMPWZ} is just the $\mathfrak{Z}_{(1,1)}$ member of this two-parametric family, while other models of \cite{MMMPWZ} are the one-parametric family $\mathfrak{Z}_{(n,n)}$.

The case considered in \cite{AMMN} corresponds to the restriction of this partition function to all $p_k=0$. Then, it becomes the sum
\be
\sum_{R}\left(\prod_{i=1}^n\xi_R(u_i)\right)S_{R}\{\bar p_k\}S_R\{g_k\}
\ee
which celebrates a series of properties: it is a$\tau$-function of the Toda lattice, it is generated by the cut-and-join rotation operators ${\cal O}$, it has multi-matrix model representation, etc. Our goal in this paper is to demonstrate that {\it all} these properties keep intact for the skew Hurwitz partition functions (\ref{Pf}). Moreover, all the properties but integrability (which is known not to survive any $\beta$-deformations) survive the $\beta$-deformation.

\bigskip

\paragraph{$w_\infty$-algebra pattern.} Another important issue, which we shortly touch in this paper, is that, from the point of view of the $W$-representations, the original technique used in \cite{China},
the partition functions $\mathfrak{Z}_{(n,n)}$ (at some special points) are generated by
combinations of operators from $w_\infty$ algebra.
The relevant for the construction are the one-parametric commutative subalgebras depicted by the blue lines
in the picture. The commutativity of operators of each line follows from the $w_\infty$-algebra relations \cite{Bakas}
\be
\left[V_{m_1,n_1}, V_{m_2,n_2}\right] = \Big((m_1-1)n_2-(m_2-1)n_2\Big) V_{m_1+m_2-2,n_1+n_2}+\ldots
\ \ \ \Longrightarrow \ \ \
\left[ {\rm ad}_{\hat F_s}^m \hat F_{s-1},  {\rm ad}_{\hat F_{s}}^{n} \hat F_{s-1}\right] = 0
\ee
for $\hat F_s = (-1)^sV_{s+1,1}$, which is nothing but the commutation of generators
\be
\left[\hat H_m^{(s)},\hat H_n^{(s)}\right] = 0 \ \ \ \ {\rm for} \ \ \
\hat H^{(s)}_m:= {\rm ad}_{\hat F_{s+1}}^m \hat F_{s}=(-1)^{m+1}V_{ms+1,m}
\ee

After the Miwa transform from time-variables to the eigenvalues of $\lambda_i$
of $N\times N$ matrices, $H_m^{(s)}$ become systems of commuting differential operators.
Not surprisingly, some of them are familiar to us:
that is, the series at $s=2$: $\hbox{ad}_{F_2}^m F_1$ (the 45-degree blue line)
turns out to be Hamiltonians of the rational Calogero-Sutherland system.
This adds to the old knowledge that the vertical line is the trigonometric Calogero-Sutherland system,
while on the horizontal line one has just commuting first-order operators $\frac{\p}{\p p_n}$.

In fact, these Calogero-Sutherland models are at the free fermion point. In order to get the Calogero-Sutherland system with a non-trivial
coupling, one needs the $\beta$-deformation. Fortunately, this whole picture survives the $\beta$-deformation, as
we explain in sec.\ref{Cal}.

\vspace{0.7cm}

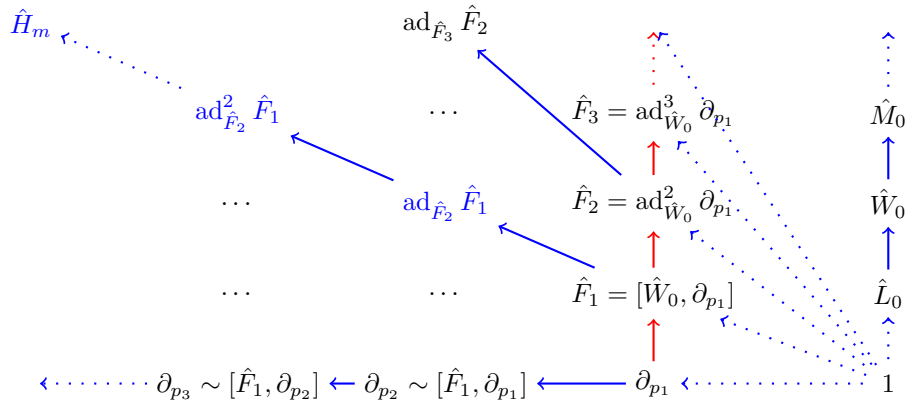
\pgfmathsetmacro{\ra}{2.3}

\begin{center}
\begin{tikzpicture}[scale=1.2]
  \foreach \x [evaluate=\x as \xval using int(\x-1)] in {3,2}
  {
  \node (mx\x) at (-1*\ra*\x,1){ $\p_{p_{\x}}\sim [\hat F_1,\p_{p_{\xval}}]$};
  }
  \node (mx1 y0) at (-\ra,1){$\p_{p_1}$};
  %\foreach \x [evaluate=\x as \xval using int(\x-1)] in {2,3}
 % {
  %\node (x\x y0) at (\ra*\x,1){};
 % }
	\node (mx4 y0) at (-4*\ra,1) {};
 % \node (x1 y0) at (\ra,1){};
  \node (x0 y1) at (0.3,1+1) {$\hat L_0$};
  \node (x0 y2) at (0.3,1+2) {$\hat W_0$};
  \node (x0 y3) at (0.3,1+3) {$\hat M_0 $};
  \node (x0 y4) at (0.3,1+4) {$ $};
  \node (mx1 y1) at (-\ra,1+1) { {$\hat F_{1}=[\hat W_0,\p_{p_1}]$}};
  \node (mx2 y1) at (-2*\ra,1+1) {\ldots};
  \node (mx3 y1) at (-3*\ra,1+1) {\ldots};
  \node (mx3 y2) at (-3*\ra,1+2) {\ldots};
  \node (mx2 y3) at (-2*\ra,1+3) {\ldots};
  \node (mx2 y2) at (-2*\ra,1+2) {${\color{blue} \operatorname{ad}_{\hat F_2}\hat F_1 } $ };
\node (mx3 y3) at (-3*\ra,1+3) {$ {\color{blue} \operatorname{ad}^2_{\hat F_2}\hat F_1 }$};
\node (mx1 y2) at (-\ra,1+2) {$\hat F_2=\operatorname{ad}_{\hat W_0}^2 \p_{p_1}$};
\node (mx1 y3) at (-\ra,1+3) {$\hat F_3=\operatorname{ad}_{\hat W_0}^3 \p_{p_1}$};
\node (mx1 y4) at (-\ra,1+4) {};
\node (mx2 y4) at (-2*\ra,1+4) {$\operatorname{ad}_{\hat F_3}\hat F_2$};
\node (mx4 y4) at (-4*\ra,1+4) {{\color{blue}$\hat H_m$}};
\node (x0 y0) at (0.3,1) {$1$};
\draw [blue,thick,->] (mx1 y1) -- (mx2 y2);\draw [blue,thick,->] (mx2 y2) -- (mx3 y3);
\draw [red,thick,->] (mx1 y0) -- (mx1 y1);\draw [red,thick,->] (mx1 y1) -- (mx1 y2);\draw [red,thick,->] (mx1 y2) -- (mx1 y3);
%\draw [blue,thick,->] (x0 y0) -- (x0 y1);
\draw [blue,thick,->] (x0 y1) -- (x0 y2);\draw [blue,thick,->] (x0 y2) -- (x0 y3);
\draw [blue,thick,->] (mx1 y2) to  (mx2 y4);
\draw [blue,thick,->] (mx1 y0) -- (mx2);
\draw [blue,thick,->] (mx2) -- (mx3);
\draw [blue,thick,loosely dotted, <-] (mx4 y4) to (mx3 y3);
\draw[red, thick, loosely dotted, ->] (mx1 y3) to (mx1 y4);
\draw[blue, thick, loosely dotted, ->] (mx3) to (mx4 y0);
\draw [blue,thick,loosely dotted, <-] (x0 y1) to (x0 y0);
\draw [blue,thick,loosely dotted, <-] (mx1 y0) to (x0 y0);
\draw [blue,thick,loosely dotted, <-] (mx1 y1) to (x0 y0);
\draw [blue,thick,loosely dotted, <-] (mx1 y2) to (x0 y0);
\draw [blue,thick,loosely dotted, <-] (mx1 y3) to (x0 y0);
\draw [blue,thick,loosely dotted, <-] (mx1 y4) to (x0 y0);
\draw [blue,thick,loosely dotted, <-] (x0 y4) to (x0 y3);
\end{tikzpicture}
\end{center}
\vspace{0.7cm}

The cut-and-join rotation operators ${\cal O}$, one of the central personages of this paper
have a spectacular interpretation in these terms:
they rotate blue lines, one into another. These operators are constructed from the generalized cut-and-join operators \cite{MMN1}.
Hence, the name. In this sense, the rational Calogero-Sutherland system is just a simple rotation of the system $\left\{\p_{p_n}\right\}$
while the trigonometric Calogero-Sutherland model is obtained from the rational one through an infinite system of rotations
trough a sequence of new integrable systems intertwined by the operators ${\cal O}$.

The paper is organized as follows. In section 2, we describe integrable properties of the generic skew Hurwitz partition function (\ref{Pf}) and its representation via generalized cut-and-join operators in parallel with \cite[secs.2-4]{AMMN}. In section 3, we concentrate on the particular case of $\mathfrak{Z}_{(1,1)}$ and derive its description as a two-matrix model depending on the external matrix and two potentials. The matrix model description can be also provided for $\mathfrak{Z}_{(n,1)}$, as we demonstrate in section 4. In section 5, we discuss the $\beta$-deformation of $\mathfrak{Z}_{(1,1)}$: the matrix model description and the $W$-representation, which turns out to be associated with the rational Calogero-Sutherland Hamiltonians. The crucial difference with the $\beta=1$ case (which corresponds to the free fermion point of the Calogero-Sutherland model) is that a description of the Hamiltonians in terms of matrix derivatives is no longer available, instead one has to use the eigenvalue variables. Section 6 contains some concluding remarks.

\section{Properties of skew Hurwitz partition functions}

\subsection{Representation via cut-and-join operators}

The partition function (\ref{Pf}) can be
realized by action of operators constructed from the commutative set of generalized cut-and-join operators $\hat W_\Delta$ \cite{MMN1}. We call these operators the cut-and-join rotation operators, they play one of the central roles in the present paper.

The generalized cut-and-join operators $\hat W_\Delta$ form a commutative set of operators, the Schur functions being their eigenfunctions \cite{MMN1}:
\be
\hat W_\Delta\ S_R=\phi_R(\Delta)\ S_R
\ee
where, for the diagram $\Delta$ containing $r$ unit cycles: $\Delta=[\tilde\Delta, 1^r]$,
\be\label{cphi}
\phi_{R,\Delta}=\left\{\begin{array}{cl}
0\ \ \ \ \ \ \ \ \ \ \ \ \ \ \ \ \ \ \ \ \ \ \ \ \ \  &|\Delta|>|R|\\
&\\
\displaystyle{(|R|-|\Delta|+r)!\over r!(|R|-|\Delta|)!}\ \phi_{R,\hat\Delta}=\displaystyle{(|R|-|\Delta|+r)!\over r!(|R|-|\Delta|)!}\
\displaystyle{\psi_{R}(\hat\Delta)\over z_{\hat\Delta}d_R}\ \ \ \ \ \ \ \ \ \ \ \ \ \ \ \ \ \ \ \ \ \ \ &|\Delta|\le |R|
\end{array}\right.
\ee
where $\hat\Delta:=[\Delta,1^{|R|-|\Delta|}]$. Now note that \cite[Eq.(61)]{MMN}
\be
\sum_\Delta \phi_R(\Delta)p_\Delta
= {S_R(p_k+\delta_{k1})\over d_R}
\ee
Now we construct the cut-and-join rotation operator as follows:
\be
\hat {\cal O}(u):=\sum_\Delta p_\Delta\cdot\hat W_\Delta,\ \ \ \hbox{with} \ p_k=u-\delta_{k,1}
\ee
Here we use the notation $p_\Delta=\prod_{i=1}^{l_\Delta}p_{\delta_i}$, where $l_\Delta$ is the length of the partition $\Delta$, and $\delta_i$'s are its parts.

Then,
\begin{equation}
    \hat {\cal O}(u) \cdot S_R\{p_k\} = \dfrac{S_R(u)}{S_R(\delta_{k,1})} S_R\{p_k\}=
    \xi_R(u)\  S_R\{p_k\}
\end{equation}
This operator was constructed earlier in \cite[Eq.(21)]{AMMN} in order to insert additional factors $\dfrac{S_R(N)}{S_R(\delta_{k,1})}$ into character expansion of the partition function, and was written there in a different form. In particular, one can rewrite it via Casimir operators \cite[sec.3]{AMMN}.

Now we can straightforwardly obtain the skew Hurwitz partition function (\ref{Pf}) by the action of these operators $\hat {\cal O}(u)$:
\be
\boxed{
\mathfrak{Z}_{(n,m)}(\{u_i\},\{v_j\}v;\bar p,g,p)=\prod_{j=1}^m\hat {\cal O}_g^{-1}(v_j)\prod_{i=1}^n\hat {\cal O}_p(u_i)\exp\left(\sum_k{(p_k+\bar p_k)g_k\over k}\right)
}
\ee

\subsection{$\mathfrak{Z}_{(n,m)}$ as a $\tau$-function of Toda lattice}

The skew Hurwitz partition function is proportional to a $\tau$-function of the Toda lattice hierarchy \cite{UT}:
\be\label{tau}
\boxed{
\tau_N(g,p)=\prod_{k=1}^N{\prod_{i=1}^n\Gamma(1-k+u_i+N)\over\prod_{j=1}^m\Gamma(1-k+v_i+N)}\cdot
\mathfrak{Z}_{(n,m)}(\{u_i+N\},\{v_j+N\};\bar p,g,p)
}
\ee
Here $N$ is the zeroth time of the hierarchy, and $\{kp_k\}$, $\{kg_k\}$ are the two infinite sets of times of the hierarchy. The third infinite set, $\{\bar p_k\}$ describes the concrete solution to the hierarchy (the point of the infinite-dimensional Grassmannian.

Let us note that, in accordance with \cite{Taka}, the sum
\be\label{Taka}
\tau_n(g,p|\xi)=\sum_{R,Q}\zeta_{R,Q}(n)S_R\{g\}S_Q\{p\}
\ee
is a $\tau$-function of the Toda lattice hierarchy iff
\be\label{Takaxi}
\zeta_{R,Q}(N)=\det_{i,j\le N}F(R_i-i+N+1,Q_j-j+N+1)
\ee
with some function $F(x,y)$, and $N$ playing the role of the zeroth time.

Hence, in order to prove that (\ref{tau}) is a Toda lattice $\tau$-function, one has to prove that
\be
\left(\prod_{k=1}^N{\prod_{i=1}^n\Gamma(1-k+u_i+N)\over\prod_{j=1}^m\Gamma(1-k+v_i+N)}\right)
{\prod_{i=1}^n\xi_R(u_i+N)\over\prod_{j=1}^m\xi_Q(v_j+N)}\cdot S_{R/Q}\{\bar p_k\}
\ee
has representation (\ref{Takaxi}). To this end, we use the Jacobi-Trudi determinant representation for the skew Schur functions,
\be
S_{R/Q}\{\bar p\}=\det_{i,j\le l_R}h_{R_i-i-Q_j+j}\{\bar p\}
\ee
where $h_k$ are the complete homogeneous symmetric polynomials, that is, $h_k=S_{[k]}$, and we put $h_k=0$ at $k<0$. Then, one obtains representation (\ref{Taka}) for (\ref{tau}) with
\be
\zeta_{R,Q}(N)=
\det_{i,j\le N}{\prod_{k=1}^n\Gamma(R_i-i+1+u_k+N)\over \prod_{l=1}^m\Gamma(Q_j-j+1+v_l+N)}\ h_{R_i-i-Q_j+j}\{\bar p\}
\ee
i.e. $\zeta_{R,Q}(N)$ is just of the form (\ref{Takaxi}) with
\be
F(x,y)={\prod_{k=1}^n\Gamma(x+u_k)\over \prod_{l=1}^m\Gamma(y+v_l)}\ h_{x-y}\{\bar p\}
\ee

\section{Two-matrix model representation of $\mathfrak{Z}_{(1,1)}$\label{2m}}

In this section, we provide the matrix model representation for the skew Hurwitz partition function $\mathfrak{Z}_{(1,1)}$. We explain that it is given by the two-matrix model
\be\label{im}
Z(N;\bar p,p,g)=\int\int_{N\times N}dXdY\exp\left(-\Tr XY+\Tr Y\Lambda+\sum_k {g_k\over k}\Tr X^k+\sum_k{\bar p_k\over k}\Tr Y^k\right)
\ee
with $p_k=\Tr\Lambda^k$.  Here the integral is understood as integration of a power series in $g_k$, $\bar p_k$ and $\Tr\Lambda^k$, and $X$ are Hermitian matrices, while $Y$ are anti-Hermitian ones.

\subsection{From matrix to time derivatives}

In order to deal with this matrix model, we need to know the action of invariant matrix derivatives, $\Tr{\p^k\over\p Y^k}$ on polynomials of $p_k=\Tr Y^k$. Note that, by comparing the actions on the Schur functions using \cite[Eq.(37)]{MMdsi} and \cite[Eq.(24)]{China}, one obtains
\begin{equation}
   \Tr\left( \frac{\p}{\p Y}\right)^k= \hat {\cal O}^{-1}(N)\left( k\dfrac{\partial}{\partial p_k}\right) \hat {\cal O}(N)
\end{equation}
Now let us use the identity\footnote{The simplest way to prove this formula is to use the Cauchy identity:
\be
\sum_RS_R\{p'_k\}S_R\left\{k{\p\over\p p_k}\right\}S_Q\{p_k\}=\exp\left(\sum_k p'_k{\p\over\p p_k}\right)S_Q\{p_k\}=
S_Q\{p_k+p'_k\}=\sum_RS_{Q/R}\{p_k\}S_R\{p'_k\}
\ee
and compare the coefficients in front of $S_R\{p'_k\}$.
}
\be
S_R\left\{k{\p\over\p p_k}\right\}S_Q\{p_k\}=S_{Q/R}\{p_k\}
\ee
in order to calculate
\be
S_R\left\{\Tr\left(\frac{\p}{\p Y}\right)^k\right\}S_Q\{\underbrace{\Tr Y^k}_{p_k}\}= \xi_Q(N)\hat {\cal O}^{-1}(N)S_R\left\{k{\p\over\p p_k}\right\}S_Q\{p_k\}=\xi_Q(N)\hat {\cal O}^{-1}(N)S_{Q/R}\{p_k\}=\nn\\
=\xi_Q(N)\hat {\cal O}^{-1}(N)\sum_PN^Q_{RP}S_P\{p_k\}=
\sum_PN^Q_{RP}\xi_Q(N)\xi_P^{-1}(N)S_P\{p_k\}
\ee
where $N^Q_{RP}$ are the Littlewood-Richardson coefficients. In particular,
\be\label{OP}
S_R\left\{\Tr\left(\frac{\p}{\p Y}\right)^k\right\}S_Q\{\underbrace{\Tr Y^k}_{p_k}\}\Big|_{Y=0}=\xi_R(N)\delta_{R,Q}
\ee

\subsection{Evaluating the matrix integral}

Now the two-matrix integral (\ref{im}) can be rewritten in the form \cite[Eq.(47)]{AMMN}
\be
Z(N;\bar p,p,g)=\sum_{R,Q}S_R\{g_k\}S_Q\{\bar p_k\}\int\int_{N\times N}dXdYS_R\{\Tr X^k\}S_Q\{\Tr Y^k\}
\exp\left(-\Tr XY+\Tr Y\Lambda\right)=\nn\\
=\sum_{R,Q}S_R\{g_k\}S_Q\{\bar p_k\}S_R\left\{\Tr\left( \frac{\p}{\p Y}\right)^k\right\}S_Q\{\Tr Y^k\}\exp\left(\Tr Y\Lambda\right)\Big|_{Y=0}
\ee
which follows from the formula of Fourier theory:
\be
\int dxdyf(x)g(y)e^{-xy}=f\Big({\p\over \p y}\Big)g(y)\Big|_{y=0}
\ee
Let us note that the combination
\be
A_{R,Q}(\Lambda):=S_R\left\{\Tr\left( \frac{\p}{\p Y}\right)^k\right\}S_Q\{\Tr Y^k\}\exp\left(\Tr Y\Lambda\right)\Big|_{Y=0}
\ee
is an invariant polynomial of $\Lambda$, i.e.
\be
A_{R,Q}(\Lambda)=\sum_P\alpha_P^{(R,Q)}S_P\{\Tr\Lambda^k\}
\ee
where $\alpha_P$ are yet unknown coefficients to be defined. Now let us apply to $A_{R,Q}(\Lambda)$ the derivative $S_P\left\{\Tr\left( \frac{\p}{\p \Lambda}\right)^k\right\}$, then put $\Lambda=0$, and use (\ref{OP}):
\be
S_R\left\{\Tr\left( \frac{\p}{\p Y}\right)^k\right\}S_Q\{\Tr Y^k\}S_P\{\Tr Y^k\}\Big|_{Y=0}=\xi_P(N)\alpha_P^{(R,Q)}
\ee
It remains to note that the l.h.s. of this equality is
\be
S_R\left\{\Tr\left( \frac{\p}{\p Y}\right)^k\right\}S_Q\{\Tr Y^k\}S_P\{\Tr Y^k\}\Big|_{Y=0}=
\sum_TN^T_{QP}S_R\left\{\Tr\left( \frac{\p}{\p Y}\right)^k\right\}S_T\{\Tr Y^k\}\Big|_{Y=0}=N^R_{QP}\xi_R(N)
\ee
Hence,
\be
\alpha_P^{(R,Q)}={\xi_R(N)\over \xi_P(N)}N^R_{QP}
\ee
and
\be\label{Zf}
\boxed{
\begin{array}{c}
Z(N;\bar p,p,g)=\sum_{R,Q,P}S_R\{g_k\}S_Q\{\bar p_k\}\alpha_P^{(R,Q)}S_P\{\Tr\Lambda^k\}=
\sum_{R,Q,P}{\xi_R(N)\over \xi_P(N)}N^R_{QP}S_R\{g_k\}S_Q\{\bar p_k\}S_P\{\Tr\Lambda^k\}=\cr
\cr
=\sum_{R,P}{\xi_R(N)\over \xi_P(N)}S_R\{g_k\}S_{R/P}\{\bar p_k\}S_P\{\Tr\Lambda^k\}=\mathfrak{Z}_{(1,1)}(N,N;\bar p,g,p)
\end{array}
}
\ee
Another derivation of this formula, which will be of use in the $\beta$-deformed case, is as follows: since the matrix integral is an invariant polynomial of $\Lambda$, one can make a replace $\Lambda\to U^{-1}\Lambda U$ with a unitary matrix $U$ and then perform an additional integration over $U$ (normalized to the volume of the unitary group $U(N)$):
\be
Z(N;\bar p,p,g)=\sum_{R,Q}S_R\{g_k\}S_Q\{\bar p_k\}S_R\left\{\Tr\left( \frac{\p}{\p Y}\right)^k\right\}S_Q\{\Tr Y^k\}\exp\left(\Tr Y\Lambda\right)\Big|_{Y=0}=\nn\\
=\sum_{R,Q}S_R\{g_k\}S_Q\{\bar p_k\}S_R\left\{\Tr\left( \frac{\p}{\p Y}\right)^k\right\}S_Q\{\Tr Y^k\}\int_{N\times N}dU\exp\left(\Tr YU^{-1}\Lambda U\right)\Big|_{Y=0}
\ee
On the other hand, this integral can be evaluated using the Itzykson-Zuber formula \cite[Eq.(4.5)]{IZ},
\be\label{IZ}
\int_{N\times N}dU\exp\left(\Tr YU^{-1}\Lambda U\right)=\sum_P{1\over\xi_P}S_P\{\Tr Y^k\}S_P\{\Lambda^k\}
\ee
immediately giving rise to (\ref{Zf}).

\section{Multi-matrix model representation of $\mathfrak{Z}_{(n,1)}$}

\subsection{Evaluating 4-matrix model $\mathfrak{Z}_{(2,1)}$}

As a natural generalization of the two-matrix model, we now consider a similar four-matrix model:
\be
&&Z^{(2)}(N;\bar p,p^{(1)},p^{(2)},g)=\int_{N_1\times N_1}dX_1dY_1\int_{N_2\times N_2}dX_2dY_2\times\nn\\
&\times&\exp\left(-\Tr X_1Y_1-\Tr X_2Y_2+\Tr Y_1\Lambda_1+\Tr Y_2\Lambda_2+\sum_k {g_k\over k}\Tr X_1^k+\sum_k{\bar p_k\over k}\Tr Y_2^k+\sum_k{\Tr Y_1^k\Tr X_2^k\over k}\right)=\nn\\
&=&\sum_{R,Q,P}S_R\{g_k\}S_Q\{\bar p_k\}S_R\left\{\Tr\left( \frac{\p}{\p Y_1}\right)^k\right\}S_P\{\Tr Y_1^k\}
S_P\left\{\Tr\left( \frac{\p}{\p Y_2}\right)^k\right\}S_Q\{\Tr Y_2^k\}\exp\left(\Tr Y_1\Lambda_1+\Tr Y_2\Lambda_2\right)\Big|_{Y_1=Y_2=0}=\nn\\
&=&\sum_{R,Q,P}S_R\{g_k\}S_Q\{\bar p_k\}A_{R,P}(\Lambda_1)A_{P,Q}(\Lambda_2)=
\sum_{R,P,Q_1,Q_2}{\xi_R\xi_P\over\xi_{Q_1}\xi_{Q_2}}S_R\{g_k\}S_{P/Q_2}\{\bar p_k\}N^R_{PQ_1}S_{Q_1}\{\Tr\Lambda_1^k\}S_{Q_2}\{\Tr\Lambda_2^k\}\nn
\ee
where $p^{(1)}_k=\Tr\Lambda_1^k$, $p^{(2)}_k=\Tr\Lambda_2^k$.

If $\Lambda_1=0$,
\be
Z^{(2)}(N_1,N_2;\bar p,p^{(2)},g)=\sum_{R,Q}{\xi_R^2\over\xi_{Q}}S_R\{g_k\}S_{R/Q}\{\bar p_k\}S_{Q}\{\Tr\Lambda_2^k\}=\mathfrak{Z}_{(2,1)}(N_1,N_2,N_2;\bar p,g,p)
\ee

\subsection{$2n$-matrix model}

Similarly, for the $2n$-matrix model, one obtains (one can compare this formula with another multi-matrix model of a similar but different type \cite{Wang:2023dey})
\be\label{2n}
&&Z^{(n)}(N_i;\bar p,\{p^{(i)}\},g)=\prod_{i=1}^n\int_{N_i\times N_i}dX_idY_i\times\nn\\
&\times&\exp\left(-\sum_{i=1}^n\Tr X_iY_i+\sum_{i=1}^n\Tr Y_i\Lambda_i+\sum_k {g_k\over k}\Tr X_1^k
+\sum_k{\bar p_k\over k}\Tr Y_n^k+\sum_{i=1}^{n-1}\sum_k{\Tr Y_i^k\Tr X_{i+1}^k\over k}\right)=\nn\\
&=&
\sum_{\{R_i,Q_i\}}S_{R_1}\{g_k\}S_{R_n/Q_n}\{\bar p_k\}\prod_{i=1}^{n-1} N^{R_i}_{R_{i+1}Q_i}\prod_{i=1}^n{\xi_{R_i}\over\xi_{Q_i}}S_{Q_i}\{\Tr\Lambda_i^k\}
\ee
and, at all $\Lambda_i=0$ but $\Lambda_n$,
\be
\boxed{
Z^{(n)}(N;\bar p,p^{(n)},g)=\sum_{R,Q}{\xi_R^n\over\xi_{Q}}S_R\{g_k\}S_{R/Q}\{\bar p_k\}S_{Q}\{\Tr\Lambda_n^k\}=\mathfrak{Z}_{(n,1)}(\{N_i\},N_n;\bar p,g,p)
}
\ee

\section{$\beta$-deformation\label{Cal}}

\subsection{The Jack polynomials}

In order to deal with the $\beta$-deformation of the matrix model (\ref{im}), we need to replace the Schur functions of sec.\ref{2m} with the Jack polynomials, and to replace correspondingly a few properties.

The orthogonality relation in this case follows from
\be\label{sJ}
J_Q\left\{{k\over\beta}{\p\over\p p_k}\right\}\ J_R\{p_k\}=||J_Q||\ J_{R/Q}\{p_k\}
\ee
where $||J_Q||$ is the norm square of the Jack polynomial,
\be
||J_R||:={\overline{G}^\beta_{R^\vee R}(0)\over G^\beta_{RR^\vee}(0)}\beta^{|R|}\ \ \ \ \ \ \
G_{R'R''}^\beta(x):=\prod_{(i,j)\in R'}\Big(x+R'_i-j+\beta(R''_j- i+1)\Big)
\ee
with the bar over the functions denoting the substitution $\beta\to\beta^{-1}$. The ratio
\be
{J_R\{N\}\over J_R\{\delta_{k,1}\}}=\prod_{i,j\in R} (N+(j-1)\beta^{-1}-i+1)
\ee

Now we again use the operator $\hat {\cal O}^\beta_N$ with the property
\be
    \hat {\cal O}^\beta(N) \cdot J_R\{p_k\} =  \xi_R^\beta(N)\  J_R\{p_k\}\nn\\
    \xi_R^\beta(N):=\prod_{i,j\in N}(N+\beta^{-1}(j-1)-i+1)
\ee
We do not need the manifest form of this operator.

Now comparing Eqs.(59) and (60) from \cite{China}, one obtains that there exists a set of commuting differential operators $\hat H_k$ such that
\begin{equation}\label{Wk}
  \hat H_k= \Big(\hat {\cal O}^\beta(N)\Big)^{-1}\left( {k\over\beta}\dfrac{\partial}{\partial p_k}\right) \hat {\cal O}^\beta(N)
\end{equation}
These operators $\hat H_n$ defined in \cite{China} do no longer have a meaning of matrix derivatives. We discuss them in detail in secs.\ref{w1}-\ref{w2}.

Now let us use the identity (\ref{sJ}) in order to obtain
\be
J_R\left\{\hat H_k\right\}J_Q\{p_k\}= \xi_Q^\beta(N)\Big(\hat {\cal O}^\beta(N)\Big)^{-1}J_R\left\{{k\over\beta}{\p\over\p p_k}\right\}J_Q\{p_k\}=\xi_Q^\beta(N)||J_Q||\Big(\hat {\cal O}^\beta(N)\Big)^{-1}J_{Q/R}\{p_k\}=\nn\\
=\xi_Q^\beta(N)||J_Q||\Big(\hat {\cal O}^\beta(N)\Big)^{-1}\sum_P\phantom{.}^{\beta^{-1}} N^{Q^\vee}_{R^\vee P^\vee}J_P\{p_k\}=
\sum_P\phantom{.}^{\beta^{-1}} N^{Q^\vee}_{R^\vee P^\vee}||J_Q||{\xi_Q^\beta(N)\over\xi_P^\beta(N)}J_P\{p_k\}
\ee
where $\phantom{.}^\beta N^Q_{RP}$ are the Littlewood-Richardson coefficients, and we used that
\be
J_{R/P}=\sum_Q\phantom{.}^{\beta^{-1}} N^{R^\vee}_{Q^\vee P^\vee}\ J_Q
\ee
Note that
\be\label{N}
\phantom{.}^{\beta^{-1}} N^{R^\vee}_{Q^\vee P^\vee}=\phantom{.}^\beta N^{R}_{Q P}{||J_R||\over ||J_Q||\ ||J_P||}
\ee
In particular,
\be\label{OP}
J_R\left\{\hat H_k\right\}J_Q\{p_k\}\Big|_{p_k=0}=\xi_R^\beta(N)||J_Q||\delta_{R,Q}
\ee

\subsection{$\beta$-deformed matrix model}

We introduce the $\beta$-deformed matrix model:
\be\label{imJ}
Z^\beta(N;\bar p,p,g)=\int\int_{N\times N}[dXdY]_\beta\exp\left(-\Tr XY+\Tr Y\Lambda+\beta\sum_k {g_k\over k}\Tr X^k+\beta\sum_k{\bar p_k\over k}\Tr Y^k\right)
\ee
where the $\beta$-deformed integration measure is {\it defined} as
\be
\boxed{
\int\int_{N\times N}[dXdY]_\beta J_R\{\Tr X^k\}J_Q\{\Tr Y^k\}
\exp\left(-\Tr XY+\Tr Y\Lambda\right)%=\nn\\
=J_R\left\{\hat H_k\right\}J_Q\{\Tr Y^k\}\exp\left(\Tr Y\Lambda\right)\Big|_{Y=0}
}
\ee
 In order to evaluate the matrix integral (\ref{imJ}), we use the same trick as before and insert an additional unitary matrix integration using the $\beta$-deformed Itzykson-Zuber formula \cite[sec.2]{MMS}
\be\label{IZb}
\int_{N\times N}[dU]_\beta\exp\left(\Tr YU^{-1}\Lambda U\right)=\sum_P{1\over\xi_P^\beta}J_P\{\Tr Y^k\}J_P\{\Lambda^k\}
\ee
Then, using the Cauchy identity for the Jack polynomials
\be
\sum_R{J_R\{p_k\}J_R\{p'_k\}\over ||J_R||}=\exp\left(\beta\sum_k{p_kp'_k\over k}\right)
\ee
one repeats the calculation of the non-deformed case and, using (\ref{OP}), immediately gets that
\be\label{imJf}
\boxed{
Z^\beta(N;\bar p,p,g)
=\sum_{R,Q,P}{\xi_R^\beta(N)\over \xi_P^\beta(N)}\phantom{.}^\beta N^R_{QP}{J_R\{g_k\}J_Q\{\bar p_k\}J_P\{\Tr\Lambda^k\}\over ||J_P||\ ||J_Q||}\stackrel{(\ref{N})}{=}\sum_{R,Q}{\xi_R^\beta(N)\over \xi_P^\beta(N)}{J_{R/Q}\{\bar p_k\}J_R\{g_k\}J_P\{\Tr\Lambda^k\}\over ||J_R||}
}
\ee
This is exactly the formula that is generated by the $W$-representation of \cite{China}.

The $\beta$-deformation of the multi-matrix model (\ref{2n}) is absolutely immediate.

\subsection{$W$-representation\label{w1}}

The $\beta$-deformed matrix model has the $W$-representation similar to that in the non-deformed case \cite{MMMPWZ}. In order to construct it, we need a set of operators $\hat H_k$ discussed above. These operators are manifestly constructed in the following way \cite{China}. First of all, we define an auxiliary operator
\be
\hat W_0&=&\dfrac{1}{2} \sum_{a, b=0}\left(\beta (a+b) p_a p_b \frac{\partial}{\partial p_{a+b}}+a b p_{a+b} \frac{\partial^2}{\partial p_a \partial p_b}\right) + {1-\beta\over 2}\sum_k (k-1)kp_k \dfrac{\partial}{\partial p_k}
\ee
which is a deformation of the cut-and-join operator \cite{GJ,MMN1}. Using this operator, we can construct a pair of another auxiliary operators
\be
\hat F_1&=&[\beta^{-1}{\p\over\p p_1},\hat W_0]=\sum_{b=0} (b+1)p_b \dfrac{\partial}{\partial p_{b+1}}\nn \\
\hat{F}_2&=&[\hat F_1,\hat W_0]=
\sum_{a,b=0} \left( \beta p_a p_b (a+b+1) \dfrac{\partial }{\partial p_{a+b+1}} + a b p_{a+b-1} \dfrac{\partial^2}{\partial p_a \partial p_b} \right) +(1-\beta) \sum_b b(b+1)p_b \dfrac{\partial }{\partial p_{b+1}}
\ee
Now we construct the whole series of operators $\hat H_k$ by the recursion relation
\be
\hat H_{k+1}={1\over k}[\hat H_k,\hat F_2]
\ee
with the initial condition $\hat H_1=\hat F_1$. The first few operators $\hat H_k$ are
\be
\hat H_1&=&\sum_{b=0} (b+1)p_b \dfrac{\partial}{\partial p_{b+1}}\nn\\
\hat{H}_2&=&[\hat H_1,\hat F_2]=
\sum_{a,b=0} \left( \beta p_a p_b (a+b+2) \dfrac{\partial }{\partial p_{a+b+2}} + a b p_{a+b-2} \dfrac{\partial^2}{\partial p_a \partial p_b} \right) +(1-\beta) \sum_{b=0} (b+1)(b+2)p_b \dfrac{\partial }{\partial p_{b+1}}\nn\\
\hat H_3&=&{1\over 2}[\hat H_2,\hat F_2]=
\sum_{a,b,c=0}\left[\beta^2(a+b+c+3)p_ap_bp_c{\p\over\p p_{a+b+c}}+abcp_{a+b+c-3}{\p^3\over\p p_a\p p_b\p p_c}\right]+\nn\\
&+&{3(1-\beta)\over 2}\sum_{a,b=0}\left[ab(a+b-2)p_{a+b-3}{\p^2\over\p p_a\p p_b}+\beta(a+b+2)(a+b+3)p_ap_b
{\p\over\p p_{a+b+3}}\right]+\nn\\
&+&{3\beta\over 2}\sum_{a,b,c,d=0}\delta_{3+a+b,c+d}p_ap_b{\p^2\over\p p_c\p p_d}+
{2\beta^2-3\beta+2\over 2}\sum_{a=0}a(a-1)(a-2)p_{a-3}{\p\over\p p_a}
\ee
and we put $p_0=N$.

With these operators one obtains for $\bar p_k=\delta_{k,2}$:
\be
Z^\beta(N;\bar p=\delta_{k,2},p,g)=e^{\hat H_2\over 2}\cdot e^{\beta\sum_k{p_kg_k\over k}}
=\sum_{R,Q,P}\beta^{|Q|\over 2}{\xi_R^\beta(N)\over \xi_P^\beta(N)}{J_{R/Q}\{\delta_{k,2}\}J_R\{g_k\}J_P\{\Tr\Lambda^k\}\over ||J_R||}
\ee
and, for $\bar p_k=\delta_{k,3}$,
\be
Z^\beta(N;\bar p=\delta_{k,3},p,g)=e^{\hat H_3\over 3}\cdot e^{\beta\sum_k{p_kg_k\over k}}
=\sum_{R,Q,P}\beta^{2|Q|\over 3}{\xi_R^\beta(N)\over \xi_P^\beta(N)}{J_{R/Q}\{\delta_{k,3}\}J_R\{g_k\}J_P\{\Tr\Lambda^k\}\over ||J_R||}
\ee
The case of generic $\bar p_k$ looks like
\be
\boxed{
Z^\beta(N;\bar p,p,g)=\exp\left(\sum_k {\beta^{1-k\over k}\bar p_k\hat H_k\over k}\right)\cdot e^{\beta\sum_k{p_kg_k\over k}}=
\sum_{R,Q,P}{\xi_R^\beta(N)\over \xi_P^\beta(N)}{J_{R/Q}\{\bar p_k\}J_R\{g_k\}J_P\{\Tr\Lambda^k\}\over ||J_R||}
}
\ee
which can be proved along the line of \cite{MMMPWZ}.

\subsection{Operators $\hat H_k$ as Calogero-Sutherland Hamiltonians\label{w2}}

Note that, for $p_k=\sum_{i=1}^N \lambda_i^k$, and when acting on symmetric functions of
$\lambda_i$,\footnote{Formal subtleties of the procedure can be found in \cite{SV}.}
the operators $\hat H_m$ are realized as
\be\label{Wn}
\hat H_1&=&\sum_i{\p\over\p\lambda_i}\nn\\
\hat H_2&=&2\beta\sum_{i\ne j}{1\over \lambda_i-\lambda_j}{\p\over\p\lambda_i}+\sum_i{\p^2\over\p\lambda_i^2}\nn\\
\hat H_3&=&3\beta^2\sum_{i\ne j\ne k}{1\over (\lambda_i-\lambda_j)(\lambda_i-\lambda_k)}{\p\over\p\lambda_i}
+3\beta\sum_{i\ne j}{1\over\lambda_i-\lambda_j}{\p^2\over\p\lambda_i^2}+\sum_i{\p^3\over\p\lambda_i^3}\nn\\
\nn\\
\ldots\nn\\
\nn\\
\hat H_n&=&\sum_{k=1}^nC^n_k\beta^{n-k}\sum_i\sum_{{I\subset [1,\dots,N]\backslash i}\atop{|I|=k}}
\prod_{j\in I}{1\over\lambda_i-\lambda_j}{\p^k\over\p\lambda_i^k}
\ee
where $C^n_k$ are the binomial coefficients. This is a set of the (mutually commuting) rational Calogero-Sutherland Hamiltonians. At $\beta=1$, it reduces to \cite[Eq.(21)]{MMM}
\be
\hat H_n=\Tr{\p^n\over\p\Lambda^n}=\sum_i\sum_{{I\subset [1,\dots,N]}\atop{|I|=n-1}}\prod_{j\in I}{1\over\lambda_i-\lambda_j}
{\p\over\p\lambda_i}
\ee
where $\Lambda$ is an $N\times N$ matrix with eigenvalues $\lambda_i$, and the operator $\hat H_n$ is understood as acting on invariant functions of $\Lambda$. Note that the sum includes the terms with poles at $i=j$, which are resolved by the L'H\^ospital's rule.

The commutativity of operators $\hat H_k$ immediately follows from their realization (\ref{Wk}). One can also consider, instead of (\ref{Wk}), the rotation \cite{MMMPWZ}
\begin{equation}
  \hat H_k^{(m)}= \prod_{i=1}^m\Big(\hat {\cal O}^\beta(u_i)\Big)^{-1}\left( {k\over\beta}\dfrac{\partial}{\partial p_k}\right)
  \prod_{i=1}^m\hat {\cal O}^\beta(u_i)
\end{equation}
which gives rise to a commutative family of Hamiltonians for each $m$. We will return to this issue elsewhere.

One can also realize $\hat H_n$ in terms of the Dunkl operators $\hat D_i$:
\be
\hat D_i={\p\over\p\lambda_i}+\beta\sum_{j\ne i}{1\over \lambda_i-\lambda_j}(1-P_{ij})
\ee
where $P_{ij}$ is the operator permuting $i$ and $j$. When acting on symmetric functions of $\lambda_i$,
\be
\hat H_k=\sum_i\hat D_i^k
\ee
Note that the standard Calogero-Sutherland Hamiltonians are obtained by the rotation:
\be
H_k^{Cal}=\Delta(\lambda)^{\beta}\ \sum_i\hat D_i^k\ \Delta(\lambda)^{-\beta}
\ee
where $\Delta(\lambda)=\prod_{i\ne j}(\lambda_i-\lambda_j)$.

Thus, we observe a surprising way of constructing the rational Calogero-Sutherland Hamiltonians: by successive commutators with $\hat F_2$ starting from $\hat H_1$, $\hat F_2$ being constructed by commutating of $\hat H_1$ with $\hat W_0$. The form of all these auxiliary and operators and Hamiltonians in terms of the matrix derivative at $\beta=1$ is
\be
\hat W_0={1\over 2}:\Tr \left(\Lambda{\p\over\p\Lambda}\right)^2: \nn\\
\hat F_1=\Tr {\p\over\p\Lambda}\nn\\
\hat F_2=\Tr \Lambda{\p^2\over\p\Lambda^2}\nn\\
\hat H_n=\Tr {\p^n\over\p\Lambda^n}
\ee
where $:\ldots :$ denotes the normal ordering, i.e. all the derivatives moved to the right. At generic $\beta$, $\hat F_1$ is still given by the same formula, while the other operators can be rewritten only in terms of the eigenvalues: the auxiliary operators are
\be
\hat W_0&=&\beta\sum_{i\ne j}{\lambda_i^2\over \lambda_i-\lambda_j}{\p\over\p\lambda_i}
+{1\over 2}\sum_i\lambda_i^2{\p^2\over\p\lambda_i^2}\nn\\
\hat F_2&=&2\beta\sum_{i\ne j}{\lambda_i\over \lambda_i-\lambda_j}{\p\over\p\lambda_i}+\sum_i\lambda_i{\p^2\over\p\lambda_i^2}
\ee
while the Hamiltonians are given by (\ref{Wn}).

Note also that $\hat W_0$ gives the trigonometric Calogero-Sutherland Hamiltonian \cite{MS,WLZZ1}. Generally, they are given by the generalized cut-and-join operators $\hat W_{[k]}$ \cite{MMN1}, and $\hat W_0=\hat W_{[2]}$. In the $w_\infty$-algebra picture of the Introduction, $\hat W_{[k]}$ are associated with the vertical line ${1\over k}V_{k+1,0}$. In particular, $\hat L_0=\hat W_{[1]}$. At the same picture,
the horizontal line is made from ${k\over\beta}{\p\over\p p_k}=V_{1,k}$.

\section{Conclusion}

In this paper, we developed the theory of skew Hurwitz partition functions. They are $\tau$-functions of the Toda lattice hierarchy of the skew hypergeometric type. Specifically, we
discussed the formalism of cut-and-join operators and of the rotation operator $\hat {\cal O}$ made from them,
and explained how to apply them to the skew Hurwitz partition functions.
This formalism allows one to substitute an explicit representation of $w_\infty$ action on the
Young diagrams \cite{MMMR,WLZZ1,China,MO} by nearly trivial manipulations with abstract operators.
We applied it to prove the equivalence of the simplest skew Hurwitz partition function to a two-matrix models in background field
(a kind of 2-matrix generalization of the generalized Kontsevich model in the character phase, \cite{GKMU}).
We also pointed out peculiarities of the $\beta$-deformation of this formalism.

We also explained the interpretation of operators $\hat {\cal O}$ as intertwiners
of commuting 1-parametric sub-algebras of $w_\infty$, which look like rotations in its pictorial representation.
After reduction to the Miwa locus (from times variables to matrix eigenvalues), these subalgebras
(the blue lines in the picture of the Introduction)
describe integrable systems, which, in the two simplest cases, are just the rational and
trigonometric Calogero-Sutherland systems having arbitrary coupling constant only after the $\beta$-deformation.
This is a pattern that deserves further analysis and better understanding.

While the $\beta$-deformation of the skew Hurwitz partition functions is immediate and preserves all the structures but integrability, the $q,t$-deformation of the picture is somewhat less straightforward, and exploits other technical means. It deserves a separate discussion.

\section*{Acknowledgements}

Our work is partly supported by the grant of the Foundation for the Advancement of Theoretical Physics ``BASIS" and by the joint grant 21-51-46010-ST-a, and by the National Natural Science Foundation of China (Nos. 11875194).

\end{document}